\documentstyle[prl,preprint,aps,epsf,floats]{revtex}
\begin{document}
\tightenlines
\newcommand{\rstev}{\mbox{$\rs = \T{1.8}$}}
\newcommand{\rssps}{\mbox{$\rs = \T{0.63}$}}
\newcommand{\XX}{\mbox{$\, \times \,$}}
\newcommand{\AP}{\mbox{${\rm \bar{p}}$}}
\newcommand{\SU}{\mbox{$<\! |S|^2 \!>$}}
\newcommand{\ET}{\mbox{$E_{T}$}}
\newcommand{\HT}{\mbox{$S_{{\rm {\sl T}}}$} }
\newcommand{\PT}{\mbox{$p_{t}$}}
\newcommand{\DP}{\mbox{$\Delta\phi$}}
\newcommand{\DR}{\mbox{$\Delta R$}}
\newcommand{\DE}{\mbox{$\Delta\eta$}}
\newcommand{\DEP}{\mbox{$\Delta\eta_{c}$}}
\newcommand{\PH}{\mbox{$\phi$}}
\newcommand{\EA}{\mbox{$\eta$} }
\newcommand{\EAJ}{\mbox{\EA(jet)}}
\newcommand{\AEA}{\mbox{$|\eta|$}}
\newcommand{\Ge}[1]{\mbox{#1 GeV}}
\newcommand{\T}[1]{\mbox{#1 TeV}}
\newcommand{\D}[1]{\mbox{$#1^{\circ}$}}
\newcommand{\x}{\cdot}
\newcommand{\ra}{\rightarrow}
\def\D0{D\O}
\def\ETmiss{{\rm {\mbox{$E\kern-0.57em\raise0.19ex\hbox{/}_{T}$}}}}
\newcommand{\mb}{\mbox{mb}}
\newcommand{\nb}{\mbox{nb}}
\newcommand{\ipb}{\mbox{$pb^{-1}$}}
\newcommand{\inb}{\mbox{$nb^{-1}$}}
\newcommand{\rs}{\mbox{$\sqrt{\rm {\sl s}}$}}
\newcommand{\fdel}{\mbox{$f(\DEP)$}}
\newcommand{\fdele}{\mbox{$f(\DEP)^{exp}$}}
\newcommand{\fgap}{\mbox{$f(\DEP\! \geq \!3)$}}
\newcommand{\fgape}{\mbox{$f(\DEP\! \geq \!3)^{exp}$}}
\newcommand{\fpyt}{\mbox{$f(\DEP\!>\!2)$}}
\newcommand{\delth}{\mbox{$\DEP\! \geq \!3$}}
\newcommand{\uplim}{\mbox{$1.1\!\times\!10^{-2}$}}
\def\simge
{\mathrel{\rlap{\raise 0.53ex \hbox{$>$}}{\lower 0.53ex \hbox{$\sim$}}}}
\def\simle
{\mathrel{\rlap{\raise 0.53ex \hbox{$<$}}{\lower 0.53ex \hbox{$\sim$}}}}


\title{The Inclusive Jet Cross Section in $\bbox{\bar{p}p}$
       Collisions at $\bbox{\sqrt{s}}$ = 1.8 TeV}

%
\author{
B.~Abbott,$^{40}$
M.~Abolins,$^{37}$
V.~Abramov,$^{15}$
B.S.~Acharya,$^{8}$
I.~Adam,$^{39}$
D.L.~Adams,$^{48}$
M.~Adams,$^{24}$
S.~Ahn,$^{23}$
H.~Aihara,$^{17}$
G.A.~Alves,$^{2}$
% ...
\\
\vskip 0.70cm
\centerline{(D\O\ Collaboration)}
\vskip 0.70cm
}
\address{
\centerline{$^{1}$Universidad de Buenos Aires, Buenos Aires, Argentina}
\centerline{$^{2}$LAFEX, Centro Brasileiro de Pesquisas F{\'\i}sicas,
                  Rio de Janeiro, Brazil}
\centerline{$^{3}$Universidade do Estado do Rio de Janeiro,
                  Rio de Janeiro, Brazil}
\centerline{$^{4}$Universidad de los Andes, Bogot\'{a}, Colombia}
\centerline{$^{5}$DAPNIA/Service de Physique des Particules, CEA, Saclay,
                  France}
\centerline{$^{6}$Panjab University, Chandigarh, India}
\centerline{$^{7}$Delhi University, Delhi, India}
\centerline{$^{8}$Tata Institute of Fundamental Research, Mumbai, India}
\centerline{$^{9}$Kyungsung University, Pusan, Korea}
\centerline{$^{10}$Seoul National University, Seoul, Korea}
\centerline{$^{11}$CINVESTAV, Mexico City, Mexico}
\centerline{$^{12}$Institute of Nuclear Physics, Krak\'ow, Poland}
\centerline{$^{13}$Institute for Theoretical and Experimental Physics,
                   Moscow, Russia}
\centerline{$^{14}$Moscow State University, Moscow, Russia}
\centerline{$^{15}$Institute for High Energy Physics, Protvino, Russia}
\centerline{$^{16}$University of Arizona, Tucson, Arizona 85721}
\centerline{$^{17}$Lawrence Berkeley National Laboratory and University of
                   California, Berkeley, California 94720}
\centerline{$^{18}$University of California, Davis, California 95616}
\centerline{$^{19}$University of California, Irvine, California 92697}
\centerline{$^{20}$University of California, Riverside, California 92521}
\centerline{$^{21}$Florida State University, Tallahassee, Florida 32306}
\centerline{$^{22}$University of Hawaii, Honolulu, Hawaii 96822}
\centerline{$^{23}$Fermi National Accelerator Laboratory, Batavia,
                   Illinois 60510}
\centerline{$^{24}$University of Illinois at Chicago, Chicago,
                   Illinois 60607}
\centerline{$^{25}$Northern Illinois University, DeKalb, Illinois 60115}
\centerline{$^{26}$Northwestern University, Evanston, Illinois 60208}
\centerline{$^{27}$Indiana University, Bloomington, Indiana 47405}
\centerline{$^{28}$University of Notre Dame, Notre Dame, Indiana 46556}
\centerline{$^{29}$Purdue University, West Lafayette, Indiana 47907}
\centerline{$^{30}$Iowa State University, Ames, Iowa 50011}
\centerline{$^{31}$University of Kansas, Lawrence, Kansas 66045}
\centerline{$^{32}$Louisiana Tech University, Ruston, Louisiana 71272}
\centerline{$^{33}$University of Maryland, College Park, Maryland 20742}
\centerline{$^{34}$Boston University, Boston, Massachusetts 02215}
\centerline{$^{35}$Northeastern University, Boston, Massachusetts 02115}
\centerline{$^{36}$University of Michigan, Ann Arbor, Michigan 48109}
\centerline{$^{37}$Michigan State University, East Lansing, Michigan 48824}
\centerline{$^{38}$University of Nebraska, Lincoln, Nebraska 68588}
\centerline{$^{39}$Columbia University, New York, New York 10027}
\centerline{$^{40}$New York University, New York, New York 10003}
\centerline{$^{41}$University of Rochester, Rochester, New York 14627}
\centerline{$^{42}$State University of New York, Stony Brook,
                   New York 11794}
\centerline{$^{43}$Brookhaven National Laboratory, Upton, New York 11973}
\centerline{$^{44}$University of Oklahoma, Norman, Oklahoma 73019}
\centerline{$^{45}$Brown University, Providence, Rhode Island 02912}
\centerline{$^{46}$University of Texas, Arlington, Texas 76019}
\centerline{$^{47}$Texas A\&M University, College Station, Texas 77843}
\centerline{$^{48}$Rice University, Houston, Texas 77005}
}

\maketitle
\begin{abstract}
We have made a precise measurement of the central inclusive jet
cross section at \rstev.  The measurement is based on an integrated
luminosity of 92 \ipb~collected at the Fermilab Tevatron
{\sl p}${\rm \bar{\sl {p}}}$ Collider with the D\O\ detector.  
The cross section,
reported as a function of jet transverse energy ($ \ET \geq \Ge{60}$)
in the pseudorapidity interval $ \AEA \leq 0.5 $, is in good agreement
with predictions from next-to-leading order quantum chromodynamics.
\end{abstract}


Within the framework of quantum chromodynamics (QCD), 
inelastic scattering between
a proton and an antiproton can be described as an elastic collision
between a single proton constituent and a single antiproton constituent.
These constituents are often referred to as partons.  After the collision,
the outgoing partons manifest themselves as localized streams of particles
or ``jets''.  Predictions for the inclusive jet cross section are given by the
folding of parton scattering cross sections with experimentally
determined parton distribution functions (pdf's).  These predictions
have recently improved with next-to-leading order (NLO) QCD scattering
calculations [1-3] and new, accurately measured pdf's [4,5].
We measure the cross section for the production of jets as a function
of the jet energy in the plane transverse to the incident beams, \ET.
The measurement is based on an integrated luminosity of
92 \ipb~of {\sl p}${\rm \bar{\sl {p}}}$ collisions collected with the
\D0 detector \cite{D0detector} at the Fermilab Tevatron Collider.
Measurements of inclusive jet production with smaller integrated luminosity
have been performed previously by the UA2 and CDF collaborations [7,8].
The cross section measurement presented here allows a stringent test of QCD,
with a total uncertainty substantially reduced relative to
previous results.

Jet detection in the \D0 detector utilizes primarily the uranium-liquid argon
calorimeters which have full coverage for pseudorapidity $|\eta| \leq 4.1$
($\eta = -{\rm ln}({\rm tan}(\theta/2))$, where $\theta$ is the polar angle
relative to the proton beam).   The calorimeter is segmented into towers of
$\DE \XX \DP = 0.1 \XX 0.1$, where $\phi$ is the azimuthal angle.

Initial event selection occurred in two hardware trigger stages and a
software stage.  The first hardware trigger selected an
inelastic {\sl p}${\rm \bar{\sl {p}}}$ collision -- indicated by
signals from the trigger hodoscopes
located near the beams on either side of the interaction region.
The next stage required transverse energy above a preset threshold
in calorimeter trigger tiles of
$\DE \XX \DP = 0.8 \XX 1.6$. Selected events
were digitized and sent to an array of processors.  Jet candidates
were then reconstructed with a cone algorithm and the
entire event recorded if any jet $\ET$ exceeded a specified threshold.
For software jet thresholds of 30, 50, 85, and \Ge{115},
integrated luminosities of 0.34, 4.6, 55, and 92
\ipb, respectively, were accumulated during a 1994--1995 data run.
The two lowest $\ET$ triggers required a single
proton-antiproton interaction within the beam crossing as signaled
by timing information in the trigger hodoscopes.

Jets were reconstructed offline using an iterative fixed-cone algorithm
with a cone
radius of $\cal{R}$=0.7 in $\eta$-$\phi$ space \cite{D0shape}.
Background from isolated noisy calorimeter cells and
accelerator beam losses which mimicked jets were eliminated with quality
cuts~\cite{Daniel}. Background events from cosmic ray bremsstrahlung or 
misvertexed
events were eliminated by requiring the missing transverse energy
in each event to be less than the larger of \Ge{30} or
$0.3E_{T}^{\rm max}$, where $E_{T}^{\rm max}$ is the \ET~of the leading jet.
Residual jet contamination is less than $1\%$ at all \ET, based
on event simulations with superimposed calorimeter noise distributions and on
visual scanning of jet candidates with \ET~greater than 350 GeV.
The jet selection efficiency for $\AEA \leq$ 0.7 has been measured
as a function of jet $\ET$ and found to be (97$\pm$1)\% below \Ge{250} and
(95$\pm$2)\% at \Ge{400}.

At high instantaneous luminosity, more than one interaction
in a single beam crossing is probable ( $\sim$~20\% for this data set).
The event vertex was reconstructed using data from the central
tracking system.
For events with multiple vertices, the two vertices with the largest number
of tracks were retained.  Due to fluctuations of jet charged-particle 
multiplicity,
an additional parameter was used to select the vertex.
If an event had more than one vertex, the quantity
\HT = $\mid\!\!\Sigma \vec{E}_{\rm {\sl T}}^{\rm jet}\!\!\mid$
was calculated for both vertices.
The vertex with the smaller \HT was selected as the event vertex and
used to calculate jet $\ET$ and $\EA$.
The selected vertex was required to be within 50 cm of the
detector center.  This last requirement retained (90$\pm$1)\% of the events,
independent of jet \ET.

The transverse energy of each jet was corrected for the
underlying event, additional interactions, noise from uranium decay,
the fraction of particle energy showered outside of the
reconstruction cone, detector
uniformity, and detector hadronic response.  A complete discussion of
the jet energy scale calibration can be found in Ref.~\cite{ESCALE}.
For $\AEA \leq 0.5$, the mean total correction factor for jet energy is
1.154$\pm$0.017 [1.118$\pm$0.023] at \Ge{100} [\Ge{400}].

The inclusive jet cross section was computed in contiguous ranges of $\ET$
using data from the four trigger sets.  The spectrum includes data
from the 30 GeV trigger between 60--\Ge{90}, from the 50 GeV
\ET~trigger between 90--\Ge{130}, then from the 85 GeV
trigger between 130--\Ge{170}, and
above \Ge{170} from the 115 GeV trigger.  The single interaction
requirement on the two lowest-$\ET$ triggers introduced an
inefficiency which was corrected by matching the 50 GeV trigger
cross section to the 85 GeV trigger cross section above 130 GeV,
where both triggers are fully efficient. This introduces an
additional 1.1\% luminosity uncertainty to the 50 GeV trigger set.
A similar matching between the lowest-$\ET$ trigger and the
50 GeV trigger introduces another 1.4\% uncertainty for the
lower set, which is added in
quadrature to the 1.1\% matching uncertainty.

The steep $\ET$ spectrum is distorted by jet energy resolution.
At all $\ET$, the resolution
(measured by balancing $\ET$ in jet events) is well described by a gaussian
distribution; at \Ge{100} the standard deviation is 7 GeV.
The distortion was corrected by assuming an ansatz function
$(A \ET ^{-B}) \cdot (1 - 2\ET/\sqrt{s} )^{C}$,
smearing it with the measured resolution, and comparing the smeared
result with the measured cross section.  The procedure was repeated
by varying parameters $A, B,$ and $C$ until the best fit was found between
the observed cross section and the smeared trial spectrum.  The ratio of the
initial ansatz to the smeared ansatz was used to correct the cross section
on a bin-by-bin basis~\cite{Mrinmoy}. The resolution correction reduces
the observed cross section by (13$\pm$3)\% [(8$\pm$2)\%] at \Ge{60} [\Ge{400}].

\begin{figure}
  \epsfxsize=6.0in
  \leavevmode\epsffile{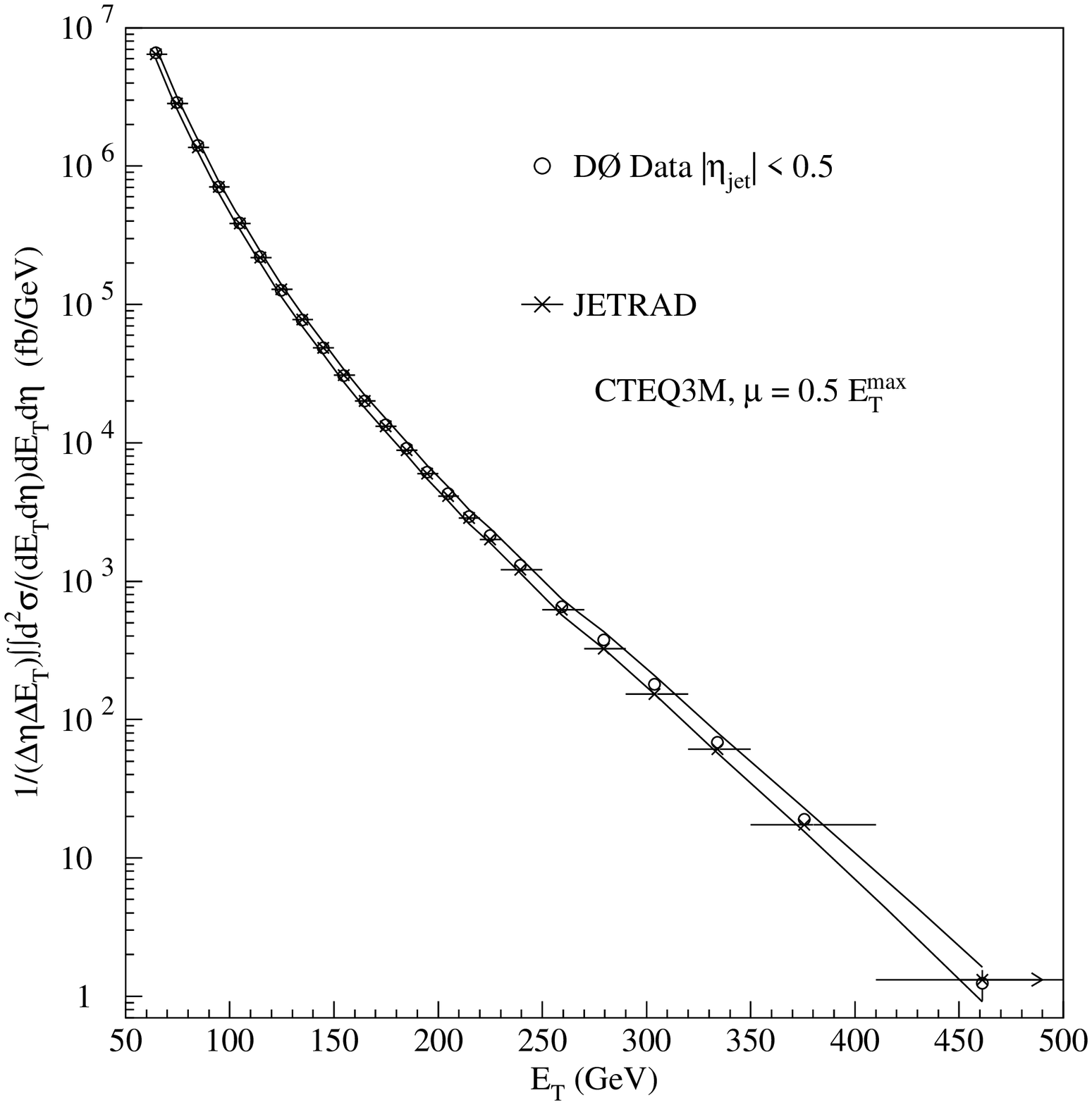}
  \caption[]{The $ \AEA \leq 0.5 $
    inclusive cross section.  Statistical uncertainties are invisible on this
    scale.  The solid curves represent the $\pm 1\sigma$ systematic
    uncertainty band.}
  \label{Fig_1}
\end{figure}

The resulting inclusive jet cross section for $\AEA \leq 0.5$, shown in 
Fig.~\ref{Fig_1},
has been averaged over each $\ET$ bin ($\Delta \ET$) and over the central
unit of rapidity ($\Delta \eta$=1).   This bin-averaged double differential
cross section, $\langle d^2\sigma / (d\ET d\eta) \rangle$,
was calculated as $N / (\Delta\ET \Delta\eta \epsilon \cal{L})$ where
$N$ is the total number of jets observed in the bin,
$\epsilon$ is the selection efficiency as a function of \ET,
and $\cal{L}$ the integrated luminosity associated with the trigger set.
The cross section is consistent with a preliminary measurement from a
smaller 1992--1993 data set~\cite{Daniel}.

Figure \ref{Fig_1} also shows a theoretical prediction for the cross
section from the NLO event generator {\small JETRAD} \cite{GGKtheory}.
There is good agreement over seven orders of magnitude.
Inputs to the NLO calculation are the renormalization scale $\mu$ (always
chosen to equal the factorization scale), the
parton distribution function (pdf), and the parton clustering algorithm.
For the calculation shown here,  $\mu = 0.5E_{T}^{\rm max}$ and the
pdf is CTEQ3M~\cite{CTEQ}.
Partons separated by less than $\cal{R}_{\rm{sep}}$=$1.3\cal{R}$ were
clustered if they were also within $\cal{R}$=0.7 of their $\ET$-weighted
$\eta$-$\phi$ centroid. This choice of $\cal{R}_{\rm{sep}}$ is discussed
in Ref.~\cite{D0shape}.  Variations in the predicted cross section due
to the input choices are approximately 30\% ~\cite{Bertram}.

The data in Fig.~\ref{Fig_1} have an overall luminosity
uncertainty of 6.1\%.  The data are plotted at the $\ET$ value
for which a smooth function describing the cross section is equal to
the average cross section in the bin. The band shows the total
systematic uncertainty as a function of \ET.
Listed in Table~\ref{tab:table1} are the plotted values of \ET, the $\ET$ 
ranges,
the cross section, and the statistical and systematic uncertainty.
The tabulated systematic uncertainties include jet and event selection,
unsmearing, relative luminosity, and energy scale uncertainties
added in quadrature.
The 6.1\% luminosity uncertainty is not included.

\begin{figure}
  \epsfxsize=6.0in
  \leavevmode\epsffile{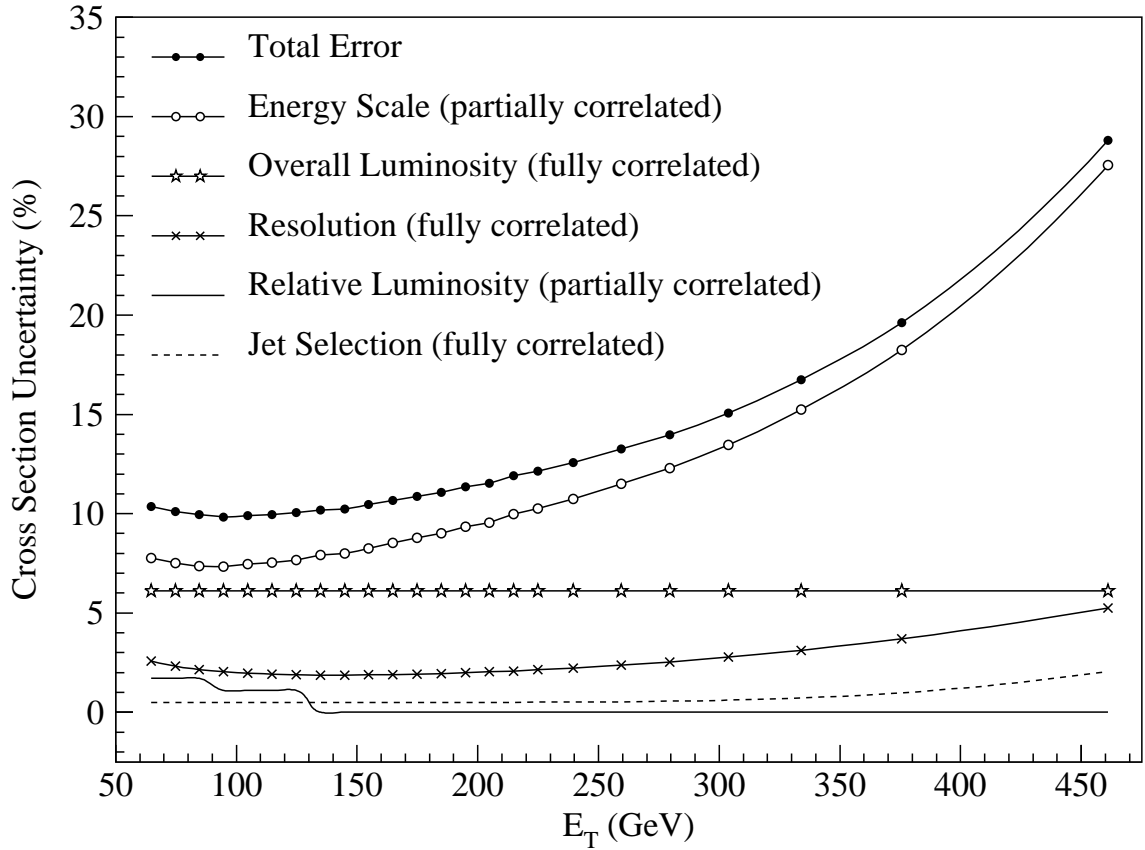}
  \caption[]{Contributions to the $ \AEA \leq 0.5 $
    cross section uncertainty plotted by component.}
  \label{Fig_2}
\end{figure}

Figure \ref{Fig_2} shows the various uncertainties for
the $\AEA \leq 0.5$ cross section.  Each curve represents the average of the
nearly symmetric upper and lower uncertainties. 
The uncertainty attributed to the
energy scale varies from 8\% at low $\ET$ to 30\% at 450 GeV.  
This contribution
dominates all other sources of uncertainty, except at low $\ET$, 
where the 6.1\% luminosity uncertainty is of comparable magnitude.

\begin{table}
\caption{  The $\AEA < 0.5$ cross section. }
\begin{tabular}{c c c c }
Plotted $E_{T}$  & Bin Range  &   Cross Sec. $\pm$ Stat.     &  Sys.  \\
  (GeV)  &  (GeV)     &        (fb/GeV)                  &  Uncer.(\%) \\ \hline
  ~64.6  & ~60~--~~70  & (6.59$\pm$0.04)$\times 10^{6}$ & ~$\pm 8$ \\
  ~74.6  & ~70~--~~80  & (2.89$\pm$0.03)$\times 10^{6}$ & ~$\pm 8$ \\
  ~84.7  & ~80~--~~90  & (1.41$\pm$0.02)$\times 10^{6}$ & ~$\pm 8$ \\
  ~94.7  & ~90~--~100  & (7.07$\pm$0.04)$\times 10^{5}$ & ~$\pm 8$ \\
  104.7  & 100~--~110  & (3.88$\pm$0.03)$\times 10^{5}$ & ~$\pm 8$ \\
  114.8  & 110~--~120  & (2.21$\pm$0.02)$\times 10^{5}$ & ~$\pm 8$ \\
  124.8  & 120~--~130  & (1.27$\pm$0.02)$\times 10^{5}$ & ~$\pm 8$ \\
  134.8  & 130~--~140  & (7.70$\pm$0.04)$\times 10^{4}$ & ~$\pm 8$ \\
  144.8  & 140~--~150  & (4.86$\pm$0.03)$\times 10^{4}$ & ~$\pm 8$ \\
  154.8  & 150~--~160  & (3.07$\pm$0.02)$\times 10^{4}$ & ~$+$9, $-$~8 \\
  164.8  & 160~--~170  & (2.00$\pm$0.02)$\times 10^{4}$ & ~$\pm 9$ \\
  174.8  & 170~--~180  & (1.34$\pm$0.01)$\times 10^{4}$ & ~$\pm 9$ \\
  184.8  & 180~--~190  & (9.12$\pm$0.10)$\times 10^{3}$ & ~$\pm 9$ \\
  194.8  & 190~--~200  & (6.15$\pm$0.09)$\times 10^{3}$ & $+$10, $-$~9 \\
  204.8  & 200~--~210  & (4.29$\pm$0.07)$\times 10^{3}$ & $\pm 10$ \\
  214.8  & 210~--~220  & (2.93$\pm$0.06)$\times 10^{3}$ & $+$11, $-$10 \\
  224.8  & 220~--~230  & (2.14$\pm$0.05)$\times 10^{3}$ & $+$11, $-$10 \\
  239.4  & 230~--~250  & (1.30$\pm$0.03)$\times 10^{3}$ & $\pm 11$ \\
  259.4  & 250~--~270  & (6.54$\pm$0.20)$\times 10^{2}$ & $+$12, $-$11 \\
  279.5  & 270~--~290  & (3.77$\pm$0.15)$\times 10^{2}$ & $+$13, $-$12 \\
  303.9  & 290~--~320  & (1.79$\pm$0.08)$\times 10^{2}$ & $+$15, $-$13 \\
  333.9  & 320~--~350  & (6.82$\pm$0.52)$\times 10^{1}$ & $+$17, $-$15 \\
  375.7  & 350~--~410  & (1.89$\pm$0.19)$\times 10^{1}$ & $+$20, $-$17 \\
  461.1  & 410~--~560  & (1.24$\pm$0.31)$\times 10^{0}$ & $+$30, $-$26 \\
\end{tabular}
\label{tab:table1}
\end{table}

The $\AEA \leq 0.5$ region provides our optimum test for departures
of data from NLO QCD.  In this region, the detector is uniformly thick 
(seven or more interaction lengths with no gaps) and both jet resolution
and calibration are precise.  Also, jet production from
the scattering of possible constituents within quarks is largest for $\eta=0$, 
relative to standard QCD predictions~\cite{Dijet}.
For comparison to Ref.~\cite{CDF}, we have also carried out a similar 
analysis in the region $0.1 \leq \AEA \leq 0.7$.

\begin{figure}
  \epsfxsize=6.0in
  \leavevmode\epsffile{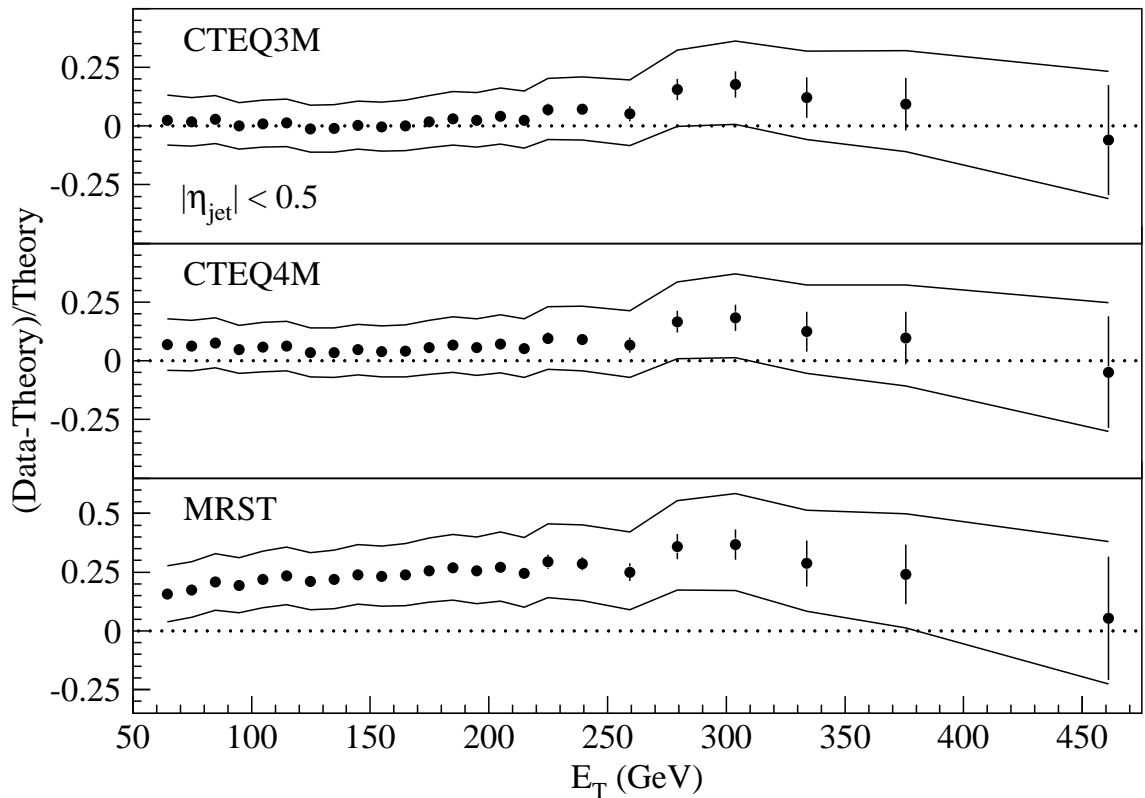}
  \caption[]{The difference between data and {\small JETRAD} QCD predictions
    normalized to predictions.  The bands represent the total experimental 
uncertainty.}
  \label{Fig_3}
\end{figure}

Figure \ref{Fig_3} shows the ratios $(D-T)/T$ for the data ($D$) and
{\small JETRAD} NLO theoretical ($T$) predictions based on the CTEQ3M, CTEQ4M
and MRST pdf's [4,5] for $\AEA \leq 0.5$.  Given the experimental
and theoretical uncertainties, the prediction is in good agreement with
the data; in particular, the data above 350 GeV show no indication of
an excess relative to QCD.

The data and theory can be compared quantitatively with a
$\chi^{2} = \sum_{i,j} (D_{i}-T_{i}) (C^{-1})_{ij} (D_{j}-T_{j})$
test incorporating the uncertainty covariance matrix
$C$. Here $D_{i}$ and $T_{i}$ represent the $i$-th data
and theory points, respectively.
The elements $C_{ij}$ are constructed by analyzing the mutual
correlation of the uncertainties in Fig.~\ref{Fig_2} at each pair of
\ET~values. As shown, the luminosity and
resolution uncertainties are quite significant at low $\ET$ and 
uncertainties in the jet energy scale
dominate at high \ET.  Since the luminosity uncertainties are fully 
correlated over all \ET, as are the resolution uncertainties,
and since the energy scale uncertainties at large $\ET$ are also
correlated, the overall systematic uncertainty is largely correlated.
Table~\ref{tab:correls} shows that the bin-to-bin correlations in the
full uncertainty for representative $\ET$ bins are greater than 40\% and 
positive.  (The full covariance matrix can be found in Ref.~\cite{matrix}.)

\begin{table}
\caption{Correlations of the total uncertainty in the cross section.}
\begin{tabular}{c c c c c c}
\ET(GeV)   & ~64.6 & 104.7 & 204.8 & 303.9 & 461.1 \\ \hline
~64.6 & 1.00  & 0.96  & 0.85  & 0.71  & 0.40  \\
104.7 & 0.96  & 1.00  & 0.92  & 0.79  & 0.46  \\
204.8 & 0.85  & 0.92  & 1.00  & 0.91  & 0.61  \\
303.9 & 0.71  & 0.79  & 0.91  & 1.00  & 0.67  \\
461.1 & 0.40  & 0.46  & 0.61  & 0.67  & 1.00  \\
\end{tabular}
\label{tab:correls}
\end{table}

Table~\ref{tab:table2} lists $\chi^{2}$ values for several {\small JETRAD}
predictions incorporating various parton distribution functions [4,5].  Each
comparison has 24 degrees of freedom.  The {\small JETRAD} predictions have
been fit to a smooth function of \ET.  All five predictions describe
the $\AEA \leq 0.5$ cross section very well (the probabilities for
$\chi^{2}$ to exceed the listed values are between 47 and 90\%).
The $0.1 \leq \AEA \leq 0.7$ cross section is also well described 
(probabilities between 24 and 72\%). 
Comparisons for $\AEA \leq 0.5$ with {\small EKS} \cite{EKStheory}
calculations using CTEQ3M, $\cal{R}_{\rm{sep}}$=$1.3\cal{R}$,
and with renormalization scales $\mu = cE_{T}^{\rm max}$ or
$cE_{T}^{\rm jet}$, where $c$ varies, are shown in Table~\ref{tab:table3}.
These calculations also describe the data very well 
(better than 57\% probability) at all renormalization scales.

\begin{figure}
  \epsfxsize=6.0in
  \leavevmode\epsffile{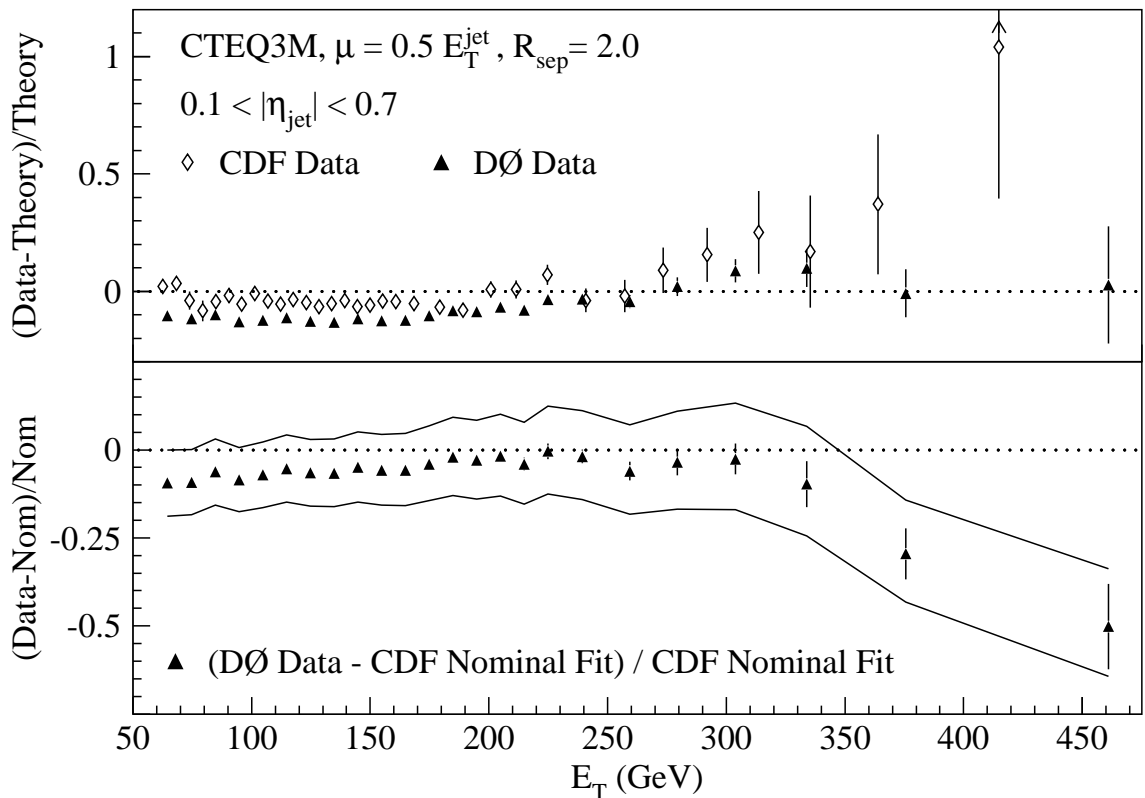}
  \caption[]{Top: Normalized comparisons of our data to {\small EKS} 
and of the data in Ref.~\cite{CDF} to {\small EKS}.  Bottom:  
Difference between our data and smoothed results of Ref.~\cite{CDF}
normalized to the latter.  The band represents the uncertainty on our data.}
  \label{Fig_4}
\end{figure}

The top panel in Fig.~\ref{Fig_4} shows $(D-T)/T$ for our data in the
$0.1 \leq \AEA \leq 0.7$ region relative to an {\small EKS} calculation 
using the CTEQ3M pdf,
$\mu=0.5E_{T}^{\rm jet}$, and  $\cal{R}_{\rm{sep}}$=$2.0\cal{R}$. (The
tabulated data can be found in Ref.~\cite{matrix}.) Also shown are
the data of Ref.~\cite{CDF} relative to the same {\small EKS} prediction.  
For this
rapidity region, we have carried out a $\chi^{2}$ comparison between our data
and the nominal curve describing the central values of the 
data of Ref.~\cite{CDF}.
Comparing our data to the nominal curve, as though it were theory, we obtain a
$\chi^{2}$ of 63.2 for 24 degrees of freedom (probability of $0.002$\%).
Thus our data cannot be described with this parameterization.  
As illustrated in the bottom panel of Fig.~\ref{Fig_4}, 
our data and the curve differ at low and high
\ET; such differences cannot be accommodated by the highly 
correlated uncertainties of our data.  If we include the systematic 
uncertainties of the data of Ref.~\cite{CDF}
in the covariance matrix, the $\chi^{2}$ is reduced to 24.7 (probability of
46\%).

\begin{table}
\caption{ $\chi^{2}$ comparisons between {\small JETRAD} and $\AEA \leq 0.5 $
and $0.1 \leq \AEA \leq 0.7 $ data for $\mu = 0.5E_{T}^{\rm max}$,
$\cal{R}_{\rm{sep}}$=$1.3\cal{R}$, and various
pdfs.  There are 24 degrees of freedom. }
\begin{tabular}{c r r}
pdf &  $\AEA \leq 0.5 $ &  $0.1 \leq \AEA \leq 0.7 $ \\ \hline
  CTEQ3M     &  23.9     &  28.4    \\
  CTEQ4M     &  17.6     &  23.3    \\
  CTEQ4HJ    &  15.7     &  20.5    \\
  MRSA\'     &  20.0     &  27.8    \\
  MRST       &  17.0     &  19.5    \\
\end{tabular}
\label{tab:table2}
\end{table}

\begin{table}
\caption{ $\chi^{2}$ comparisons between {\small EKS} and $\AEA \leq 0.5$ data
for CTEQ3M, $\cal{R}_{\rm{sep}}$=$1.3\cal{R}$, and various
renormalization scales given by $cE_{T}^{\rm max}$ or $cE_{T}^{\rm jet}$.
There are 24 degrees of freedom.}
\begin{tabular}{c r r}
  $c$ &  $\chi^{2}$ ($E_{T}^{\rm max}$)& $\chi^{2}$ ($E_{T}^{\rm jet}$) \\
  \hline
 0.25 & 14.8 & 19.8 \\
 0.50 & 19.4 & 22.2 \\
 1.00 & 16.8 & 18.1 \\
\end{tabular}
\label{tab:table3}
\end{table}

In conclusion, we have made the most precise measurement to date
of the inclusive jet cross section for $\ET \geq \Ge{60}$.  QCD
predictions are in good agreement with the observed cross section
for standard parton distribution functions and different renormalization 
scales.  This is consistent with our previous measurements of dijet 
angular distributions \cite{Dijet}, which are also in good agreement with 
QCD and show no indication of physics beyond the Standard Model.

%
We thank the staffs at Fermilab and collaborating institutions for their
contributions to this work, and acknowledge support from the
Department of Energy and National Science Foundation (U.S.A.),
Commissariat  \` a L'Energie Atomique (France),
Ministry for Science and Technology and Ministry for Atomic
   Energy (Russia),
CAPES and CNPq (Brazil),
Departments of Atomic Energy and Science and Education (India),
Colciencias (Colombia),
CONACyT (Mexico),
Ministry of Education and KOSEF (Korea),
and CONICET and UBACyT (Argentina).
We thank W.~T.~Giele, E.~W.~N.~Glover, and D.~A.~Kosower for
help with {\small JETRAD} and B.~Flaugher for help with data
comparisons.

\end{document}